# Fostering Inclusive Virtual Reality Environments: Discussing Strategies for Promoting Group Dynamics and Mitigating Harassment

NILOOFAR SAYADI, University of Notre Dame, USA

DIEGO GÓMEZ-ZARÁ, University of Notre Dame, USA

## 1 INTRODUCTION

The rapid evolution of social Virtual Reality (VR) platforms has significantly enhanced the way users interact and socialize in digital spaces, offering immersive experiences that closely mimic real-world interactions [1]. However, this technological advancement has brought new challenges, particularly in ensuring safety and preventing harassment [11]. Unlike traditional social media platforms, the immersive nature of social VR applications can intensify the impact of harassment, affecting users' emotions, experiences, and reactions at both mental and physical levels. [2, 9].

Group dynamics can play a pivotal role in both preventing and mitigating harassment [9]. Literature in group dynamics has provided insights into fostering and nurturing communities, with special emphasis on how enabling leadership, coordination, and cohesion among individuals can enable inclusive and safer social spaces [4]. For this workshop, we propose to discuss group-centric approaches to address harassment in social VR. We will discuss group strategies such as matching, interactions, and reporting mechanisms aimed at promoting safer and more supportive social VR spaces. By leveraging the social structures among users, we aim to empower users and communities to collectively counteract harassment and ensure a positive social experience for all users.

## 2 STRATEGIES TO COMBAT HARASSMENT WITHIN VR TEAMS

To address harassment in VR environments, we propose to discuss several strategies inspired by enabling group dynamics.

### 2.1 Facilitating Safe Spaces Through Group Membership.

Based on self-categorization theories [7, 8], which suggests that individuals categorize themselves and others into groups based on shared characteristics or interests, we propose to enable group matching strategies in social VR platforms that can support users from experiencing harassment. When new users enter a virtual space, they often lack established social connections or an understanding of community norms, making them vulnerable to harassment [6, 9]. By implementing a group matching strategy that adds new users to groups with similar interests, backgrounds, or values, users can build a sense of belonging and establish a supportive network that deters potential harassers [9]. Moreover, in the presence of strangers, social VR platforms can orient users to find established members of the community, who can be approachable and willing to listen and help. Users can visualize cues and recommendations that can guide their interactions in order to assist new introductions. Orienting users from the beginning can quickly

Authors' addresses: Niloofar Sayadi, University of Notre Dame, USA, nsayadi2@nd.edu; Diego Gómez-Zará, University of Notre Dame, USA, dgomezara@nd.edu.

identify support groups. In doing so, users can collectively combat harassment or negative social interactions together [10].

Another potential strategy is designing "group bubbles" to provide additional protection for users. Like personal "bubbles" that prevent others from invading a user's personal space, group bubbles can function as private spaces where access is strictly controlled and limited to group members [3]. Although personal bubbles can effectively prevent unwanted virtual interactions on an individual level, they may isolate users from group interactions, making it harder to form connections and participate in group activities. By creating a virtual space exclusive to the group, members can have more control over not only their surroundings but also who they can interact with when privacy is necessary. Additionally, these group bubbles can include customizable settings to provide group control over the size of their bubble, individuals (either friends, moderators, or similar users) who could see be part of it [10].

### 2.2 Reporting mechanisms

Previous studies make a compelling argument for the need for robust reporting systems to protect users from identified risks and harms [5, 12]. Given this knowledge, we aim to discuss new designs that promote real-time and easy-to-use reporting systems that flag harassment quickly and effectively [11, 13]. For example, a potential solution is implementing an alert system that notifies online users or those near the user about a harassment situation. Enabling alerts would increase awareness and accountability in social spaces, enabling users to quickly inform their group when a harasser is present. The alert could include information about the location of the incident within the virtual space and the identity of the harasser if known. Group members receiving the alert would then be aware of the danger and take appropriate actions, such as coming to the aid of the targeted user, reporting the harasser to platform moderators, or collectively blocking the harasser from the group. As a result, users could collectively recognize and respond to the threat. Such a system could act as a rapid response mechanism, fostering a sense of security and collective vigilance within the group.

## 3 CONCLUSION

The immersive and embodied nature of social VR introduces new forms of harassment that require innovative solutions. Based on our expertise in groups, which encompasses how individuals form groups and build support, we aim to discuss how social VR platforms can employ group dynamics to significantly reduce the risk of harassment and promote inclusion among different users. As social interactions can be curated, designed, and shaped by the use of space and immersion, our goal is to reflect on the interplay of group dynamics and social VR to protect users from harm, discrimination, and harassment. Strategies such as group matching or group bubbles can help users avoid harassment and emulate social actions and support networks that prevail in face-to-face contexts. Therefore, we aim to contribute to this collective discussion and work for new solutions in the space of social VR.

Fostering Inclusive Virtual Reality Environments 3## REFERENCES

[1] Dane Acena and Guo Freeman. 2021. "in my safe space": Social support for lgbtq users in social virtual reality. In *Extended abstracts of the 2021 CHI conference on human factors in computing systems*. 1–6.

[2] Lindsay Blackwell, Nicole Ellison, Natasha Elliott-Deflo, and Raz Schwartz. 2019. Harassment in social virtual reality: Challenges for platform governance. *Proceedings of the ACM on Human-Computer Interaction* 3, CSCW (2019), 1–25.

[3] Andrea Bönsch, Sina Radke, Heiko Overath, Laura M Asché, Jonathan Wendt, Tom Vierjahn, Ute Habel, and Torsten W Kuhlen. 2018. Social VR: How personal space is affected by virtual agents' emotions. In *2018 IEEE Conference on Virtual Reality and 3D User Interfaces (VR)*. IEEE, 199–206.

[4] Amy C Edmondson and Zhike Lei. 2014. Psychological safety: The history, renaissance, and future of an interpersonal construct. *Annu. Rev. Organ. Psychol. Organ. Behav.* 1, 1 (2014), 23–43.

[5] Guo Freeman, Samaneh Zamanifard, Divine Maloney, and Dane Acena. 2022. Disturbing the peace: Experiencing and mitigating emerging harassment in social virtual reality. *Proceedings of the ACM on Human-Computer Interaction* 6, CSCW1 (2022), 1–30.

[6] Aurora Harley. 2020. Similarity Principle in Visual Design.

[7] Matthew J. Hornsey. 2008. Social Identity Theory and Self-categorization Theory: A Historical Review. *Social and Personality Psychology Compass* 2, 1 (2008), 204–222. https://doi.org/10.1111/j.1751-9004.2007.00066.x arXiv:https://compass.onlinelibrary.wiley.com/doi/pdf/10.1111/j.1751-9004.2007.00066.x

[8] Felipe León. 2023. Being one of us: we-identities and self-categorization theory. *Phenomenology and the Cognitive Sciences* (2023), 1–25.

[9] Jinghuai Lin and Marc Erich Latoschik. 2022. Digital body, identity and privacy in social virtual reality: A systematic review. *Frontiers in Virtual Reality* 3 (2022), 974652.

[10] Divine Maloney and Guo Freeman. 2020. Falling asleep together: What makes activities in social virtual reality meaningful to users. In *Proceedings of the Annual Symposium on Computer-Human Interaction in Play*. 510–521.

[11] Regan L Mandryk, Julian Frommel, Nitesh Goyal, Guo Freeman, Cliff Lampe, Sarah Vieweg, and Donghee Yvette Wohn. 2023. Combating Toxicity, Harassment, and Abuse in Online Social Spaces: A Workshop at CHI 2023. In *Extended Abstracts of the 2023 CHI Conference on Human Factors in Computing Systems*. 1–7.

[12] Joseph O'Hagan, Florian Mathis, and Mark McGill. 2023. User Reviews as a Reporting Mechanism for Emergent Issues Within Social VR Communities. In *Proceedings of the 22nd International Conference on Mobile and Ubiquitous Multimedia*. 236–243.

[13] Sarita Schoenebeck, Oliver L Haimson, and Lisa Nakamura. 2021. Drawing from justice theories to support targets of online harassment. *new media & society* 23, 5 (2021), 1278–1300.Manuscript submitted to ACM